\documentclass[fleqn,usenatbib,onecolumn]{mnras}
\usepackage{lmodern}
\usepackage{newtxmath}
\usepackage{newtxtext,newtxmath}

\usepackage[switch]{lineno} 

\usepackage[T1]{fontenc}

\DeclareRobustCommand{\VAN}[3]{#2}
\let\VANthebibliography\thebibliography
\def\thebibliography{\DeclareRobustCommand{\VAN}[3]{##3}\VANthebibliography}

\usepackage{graphicx}	
\usepackage{amsmath}	
\usepackage{enumitem}
\setdescription{font=\sffamily\bfseries, style=nextline, leftmargin=2em}

\title[Are jet speeds governed by accretion modes?]{Are jet speeds governed by accretion modes?}

\author[H. Tu et al.]{
Hong Tu,$^{1,2}$\thanks{E-mail: tuhong@shnu.edu.cn (HT)}
Xinwu Cao,$^{3,4}$\thanks{E-mail: xwcao@zju.edu.cn (XC)}
and Andrzej A. Zdziarski$^{5}$
\\
$^{1}$Department of Physics, Shanghai Normal University, 100 Guilin Road, Shanghai, 200234, China\\
$^{2}$Shanghai Key Lab for Astrophysics, Shanghai Normal University, 100 Guilin Road, Shanghai, 200234, China\\
$^{3}$Institute for Astronomy, School of Physics, 866 Yuhangtang Road, Hangzhou, 310058, Zhejiang, China\\
$^{4}$Center for Cosmology and Computational Astrophysics, School of Physics, 866, Yuhangtang Road, Hangzhou, 310058, Zhejiang, China\\
$^{5}$Nicolaus Copernicus Astronomical Center, Polish Academy of Sciences, Bartycka 18, Warszawa, PL-00-716, Poland}

\date{Accepted 2026 September 02. Received 2026 September 02; in original form 2026 May 11}

\pubyear{\the\year{}}

\begin{document}
\label{firstpage}
\pagerange{\pageref{firstpage}--\pageref{lastpage}}
\maketitle

\begin{abstract}
Relativistic jets are observed in both stellar-mass black hole X-ray binaries (BHXRBs) and active galactic nuclei (AGNs), yet their bulk Lorentz factors differ systematically---those in black hole X-ray binaries are typically below $\sim2$, whereas AGN jets can reach $\sim50$. The origin of this discrepancy remains unclear. Searching the literature, we compile a sample of 333 AGNs with well-measured jet component motions, consisting of 270 quasars, 47 BL Lac objects, 10 FR I, and 6 FR II galaxies. We find that quasars/FR\,IIs exhibit minimal bulk Lorentz factors ranging from $\sim$1.0 to 41.5, with a mean of 11.5 (median 9.5). In contrast, BL Lac objects/FR\,Is show $\Gamma_{\rm jet}\sim$1.0--21.9, averaging 4.2 (median 1.5). These values, particularly the median, closely resemble those of BHXRBs, implying a strong correlation between jet speed and accretion mode. The Lorentz factor of a magnetically driven jet is mainly determined by the ratio of the magnetic pressure to rest mass energy density at the jet base. In BL Lacs/FRIs/BHXRBs, the field is maintained by the advection-dominated accretion flow (ADAF), and the gas at the ADAF surface is magnetically driven into the jets. In quasars/FRIIs, the field is maintained by the disc, while the jet base is connected to the corona. Our model calculations show that the disc field is always much stronger than that of the ADAF, and therefore leads to a larger $\Gamma_{\rm jet}$, which can explain the systematic difference in $\Gamma_{\rm jet}$ between these two types of sources.  
\end{abstract}



\section{Introduction}\label{sec:intro}

Different states and rapid transitions between them in black hole X-ray binaries (BHXRBs) provide useful clues for testing accretion disc theories. It is believed that the states of BHXRBs correspond to different accretion modes \citep{1997ApJ...489..865E}. In the high state, a geometrically thin accretion disc extends to the BH's innermost stable circular orbit (ISCO). In the low/intermediate states, the thin disc is truncated at a large distance from the BH, and a geometrically thick, hot accretion flow replaces the thin disc in the inner region surrounding the BH \citep{1997ApJ...489..865E, 2004MNRAS.351..791Z, 2010LNP...794...53B, 2022abn..book.....C}. 

Direct observation of such state transitions in massive BH counterparts seems unlikely, as the timescales for state transitions in BHXRBs are too long for active galactic nuclei (AGNs). Although a small fraction of changing-look AGNs (CLAGNs) exhibit type transitions on timescales of years to decades, the physics underlying these transitions is believed to differ from that responsible for state transitions in BHXRBs \citep[see][for a recent review and the references therein]{2024SerAJ.209....1K}. Observations of different types of AGNs indeed reveal distinct accretion modes. For radio Fanaroff-Riley (FR) galaxies \citep{1974MNRAS.167P..31F}, FR II galaxies usually have broad emission lines in their spectra, whereas only narrow emission lines have been detected in FR I galaxies, implying different accretion modes in these two types of FR galaxies \citep{2016aapRv..24...10T}. \citet{2001aap...379L...1G} suggested that FR I galaxies can be separated from FR II galaxies by an Eddington ratio of $\sim 0.001-0.01$ based on a sample of FR galaxies \citep{1995ApJ...451...88B}, indicating that the accretion modes in the two types are controlled by the accretion rate. In the unification scheme for radio-loud AGNs, BL Lacertae (BL Lac) objects are believed to be FR I radio galaxies with jets aligned with our line of sight, while radio quasars correspond to FR II galaxies viewed at small angles relative to the jet orientation \citep{1995PASP..107..803U, 2016aapRv..24...10T}. A similar division in Eddington ratio has been found between BL Lac objects and quasars \citep{2009ApJ...694L.107X}, implying that radio quasars may contain radiatively efficient accretion discs, whereas radiatively inefficient, hot accretion flows may be present in BL Lac objects. Thus, BL Lac objects appear to be the massive BH counterparts of BHXRBs in low/intermediate states, whereas quasars are the counterparts of high-state BHXRBs.  

Steady relativistic jets are consistently observed in the low/hard state of BHXRBs, whereas they are typically suppressed in the high/soft state of low-mass BHXRBs \citep[e.g.,][]{1999MNRAS.308..473F, 2001ApJ...554...43C, 2003MNRAS.344...60G, 2004MNRAS.355.1105F}. However, jets appear to be present in both low-luminosity and luminous AGNs. Jets are most likely powered by tapping energy from spinning BHs via a corotating magnetic field \citep{1977MNRAS.179..433B}. In this scenario, both a strong magnetic field and a rapidly spinning BH are required to launch jets from the BH horizon. Such a strong magnetic field with a net flux near the BH must be formed through accretion from large distances. However, field advection in a conventional viscously driven thin disc is very inefficient \citep{1994MNRAS.267..235L}, whereas a weak field in the outer region can be dragged inwards by an advection-dominated accretion flow (ADAF) due to its large radial velocity \citep{2011ApJ...737...94C, 2023ApJ...944..182D}. 

The BH accretion discs are similar in both BHXRBs and AGNs. However, in BHXRBs, the mass feeding the BHs comes from the donors, whereas in AGNs, the BHs are fed by the circumnuclear interstellar medium. It has been suggested that geometrically thin accretion discs with magnetic outflows are present in luminous radio-loud AGNs because the interstellar medium provides both mass and sufficient magnetic flux to the outer disc. Most angular momentum of such a disc is removed by the outflows, and the radial velocity of the disc is significantly increased, which leads to efficient magnetic field advection through the thin disc to produce a strong field at the BH horizon that launches relativistic jets \citep{2013ApJ...765..149C, 2016ApJ...833...30C}. \citet{2019MNRAS.485.1916C} suggested that the field strength at the BH horizon is too weak to drive jets in the high state of BHXRBs, as the field of their low-mass donors is not sufficiently strong to maintain a disc-outflow system like those in luminous AGNs, which provides a natural explanation for why jets are suppressed in high-state low-mass BHXRBs.   

There are two main types of jets in XRBs: compact, steady jets are always associated with the low/hard state, while transient, discrete jets appear occasionally during transitions from the hard intermediate state to the soft state. The main difference between these two types of jets lies in their radio spectra: flat or even inverted in compact jets, whereas steep in transient jets. However, the bulk Lorentz factors are quite similar for both types of jets \citep[see][for a short review and the references therein]{2024ApJ...967L...7Z}. Most are in the range of $\lesssim 2$ \citep[e.g.,]{2001MNRAS.327.1273S,2005ApJ...632..504C,2010MNRAS.404L..21C, 2012ApJ...745..136S, 2022ApJ...935L...4Z}. A few estimates show slightly higher Lorentz factors up to $\sim 3-4$ \citep[e.g.,][]{2003ApJ...592..347W, 2022ApJ...925..189Z, 2025arXiv250411945Z}, or even higher, which have not been observed due to selection effects, as speculated by \citet{2026MNRAS.545f2102L,2026MNRAS.tmp.1410L}.   
The bulk Lorentz factors of AGN jets span a wider range, up to several tens \citep[e.g.,][]{1997AJ....114.1999S, 1998AJ....115.1357S, 2009AJ....138.1874L, 2013AJ....146..120L}, and are systematically higher than those in BHXRBs. Both AGNs and XRBs have similar central engines: a BH surrounded by an accretion disc and producing relativistic jets; however, the reason for the difference in jet speeds remains a mystery. Whether the Lorentz factors of the jets are regulated by other physical quantities of AGNs has been explored in some previous works  \citep{2009RAA.....9..293Z,2012ApJ...759..114C}.

In this work, we use a sample of radio AGNs consisting of quasars and BL objects with well-measured proper motions of components in their relativistic jets. We analyze the distributions of the minimal bulk Lorentz factors for these two source types in Sections \ref{sec:sample} and \ref{sec:static-results}. The comparison between these two distributions may shed light on the relationship between jet acceleration and the accretion disc. Because hot accretion flows are believed to be present in both BL Lac objects and BHXRBs in low/intermediate states, we suggest that comparing the Lorentz factor distribution of BL Lac objects with that of BHXRBs is more appropriate than comparing it with that of the whole sample of AGNs. Finally, we develop a model for jet formation with different accretion modes, based on the Blandford-Znajek mechanism \citep{1977MNRAS.179..433B}, in Section \ref{sec:mag-jets}. This model clearly shows how the jet speed is regulated by accretion modes and reproduces the discrepancy in the Lorentz factor distributions between quasars/FR IIs and BL objects/FR Is (or BHXRBs). The last section contains a discussion of the results.  

\section{Sample} \label{sec:sample}

Parsec-scale jet kinematics have been extensively studied in many previous works \citep[e.g.,][]{1998AJ....115.1295K, 2009AJ....138.1874L, 2017ApJ...846...98J}. {We construct our sample primarily from the MOJAVE program \citep{2021ApJ...923...67H}, which presented parsec-scale jet kinematics for 409 bright radio-loud AGNs based on 15 GHz VLBA data, tracking 1744 individual features in 382 jets over at least 5 epochs. To build a more complete and robust sample, we also searched the literature extensively and included additional sources with reliable redshift and kinematic measurements. We excluded objects with uncertain redshift determinations, particularly the BL Lac objects whose redshifts were estimated from host galaxy luminosity as a standard candle \citep{2010ApJ...712...14M}, as well as radio galaxies whose morphological classifications are ambiguous. The filtering and classification procedures are detailed below. The notes on the sources are summarized as follows.

\begin{description}

\item[Misclassified sources] Two sources in the original MOJAVE sample were misclassified as BL Lac objects but have since been reclassified as quasars based on optical spectroscopy. These are listed in Table~\ref{tab:misclassified}. In addition, the source S4~0954+65 (originally classified as BL Lac in the same sample) has been suggested to be a transitional blazar with properties intermediate between BL Lacs and FSRQs (e.g., \citealt{2023ApJ...949...39G}); we therefore exclude it from our statistical analysis to avoid ambiguity arising from its uncertain classification.

\begin{table}
\centering
\caption{Misclassified sources}
\label{tab:misclassified}
\begin{tabular}{lllll}
\hline
ID & Alias & Original Class & Corrected Class & Reference \\
\hline
0044+566 & GB6 J0047+5657 & B (BL Lac) & Q (Quasar) & \cite{2005ApJ...626...95S} \\
1656+482 & 4C +48.41 & B (BL Lac) & Q (Quasar/FSRQ) & \cite{2011ApJ...743..171A} \\
\hline
\end{tabular}
\end{table}

\item[BL Lac objects excluded (uncertain redshifts).] We compiled a BL Lac sample based on the MOJAVE catalog, supplemented with additional sources from the literature. We excluded a total of 18 BL Lac objects with uncertain redshift determinations from this sample, as detailed in Table~\ref{tab:excluded-bllac}. After removing these sources, our final sample consists of 47 BL Lac objects.

\begin{table}
\centering
\caption{BL Lac objects excluded due to uncertain redshifts}
\label{tab:excluded-bllac}
\begin{tabular}{lll}
\hline
ID & Alias & Redshift Determination \\
\hline
0106+678 & 4C +67.04 & Estimated from host galaxy (standard candle) \cite{2010ApJ...712...14M} \\
0138-097 & PKS 0139-09 & Estimated from host galaxy (standard candle) \cite{2013ApJ...764..135S} \\
0214+083 & PMN J0217+0837 & Estimated from host galaxy (standard candle) \cite{2013ApJ...764..135S} \\
0219+428 & 3C 66A & Lower limit from EBL absorption \cite{2018MNRAS.474.3162T} \\
0313+411 & IC 310 & Estimated from host galaxy \cite{2013ApJ...764..135S} \\
0420+417 & 4C +41.11 & Tentative redshift ($z=0.397$) based on a single [O III] line only \cite{2019ATel12802....1P} \\
0708+506 & GB6 J0712+5033 & Estimated from host galaxy \cite{2013ApJ...764..135S} \\
0716+714 & S5 0716+714 & Statistically inferred from group membership ($z=0.2304$) \cite{2023aap...680A..52P} \\
0829+046 & PKS 0829+046 & Derived from neighboring galaxies, not reliable \cite{1994AJ....107..494P} \\
0859+210 & NVSS J090226+205045 & Lower limit from host galaxy features ($z \ge 0.706$) \cite{2026aap...709A.107F} \\
0925+504 & GB6 J0929+5013 & Upper limit from EBL absorption \cite{2024MNRAS.527.4763D} \\
1139+160 & MG1 J114208+1547 & Estimated from host galaxy \cite{2013ApJ...764..135S} \\
1206+416 & B3 1206+416 & Estimated from host galaxy \cite{2013ApJ...764..135S} \\
1219+285 & W Comae & No reliable redshift \cite{2017ApJ...837..144P} \\
1250+532 & S4 1250+53 & Upper limit from EBL absorption \cite{2024MNRAS.527.4763D} \\
1307+121 & OP 112 & No reliable redshift \cite{2021AJ....161..196P, 2022ApJS..259...55N} \\
1413+135 & PKS 1413+135 & Only lower limit ($z>0.247$) from H I and OH absorption lines \cite{2023aap...671A..43C} \\
2023+760 & S5 2023+760 & Measured from neighbouring galaxy features \cite{2013ApJ...764..135S} \\
\hline
\end{tabular}
\end{table}

\item[Radio Galaxies Classification] We compiled a sample of radio galaxies with well-established Fanaroff-Riley (FR) classifications from the MOJAVE survey and additional literature. Table~\ref{tab:fr-all} lists the 16 sources that have unambiguous FR type I or II, along with their optical class, redshift, observed jet speed $\beta_{\rm max}$ and its uncertainty, and the corresponding references for both the kinematic measurements and the FR classification.

\begin{table}
\centering
\caption{FR classification of radio galaxies}
\label{tab:fr-all}
\begin{tabular}{llll}
\hline
ID & Alias & FR Class & Reference \\
\hline
0055+300 & NGC 315 & I  & \cite{2012MNRAS.420.2715C} \\
0305+039 & 3C 78    & I  & \cite{1999MmSAI..70..141T} \\
0309+411 & NRAO 128 & I  & \cite{2010aap...518A..10V} \\
0316+413 & 3C 84    & I  & \cite{2021MNRAS.500.4671L} \\
0430+052 & 3C 120   & I  & \cite{2015ApJ...799L..18T} \\
1228+126 & M87      & I  & \cite{1974MNRAS.167P..31F} \\
1637+826 & NGC 6251 & I  & \cite{2003aap...410..131G} \\
1638+118 & TXS 1638+118 & I  & \cite{2021MNRAS.505.5853G} \\
1222+131 & M84      & I  & \cite{2011MNRAS.417.2789L} \\
1224+135 & NGC 4261 & I  & \cite{2015MNRAS.450.1732K} \\
\hline
0007+106 & III Zw 2 & II & \cite{2008ApJ...674..111C} \\
0415+379 & 3C 111   & II & \cite{1998ASPC..144..129A} \\
1128-047 & PKS 1128-047 & II & \cite{2013Chaap..37...28X} \\
1833+326 & 3C 382   & II & \cite{2010MNRAS.401L..10T} \\
1845+797 & 3C 390.3 & II & \cite{2003AJ....126.2677Q} \\
1957+405 & Cygnus A & II & \cite{2020ApJ...891..173S} \\
\hline
\end{tabular}
\end{table}
\end{description}

Thus, our final sample for analysis comprises 333 AGNs: 270 quasars, 47 BL Lac objects, and 16 morphologically classified radio galaxies (10 FR\,Is and 6 FR\,IIs). The full sample table, including ID, alias, redshift, optical class, maximum apparent speed, its uncertainty, and reference, is available as a machine-readable table in the online supplementary material; a small portion (the first 10 sources) is shown in Table~\ref{tab:AGN_kinematics_sample} for illustration. Using the maximal proper motion data derived from multi-epoch VLBI observations, the minimal bulk Lorentz factors for the fastest-moving components of the jets are derived as
\begin{equation}
    \Gamma_{\rm min}=(1+\beta_{\rm app}^2)^{1/2}. \label{gamma_min}
\end{equation}

\begin{table}
\centering
\caption{Example of the sample table (first 10 sources).}
\label{tab:AGN_kinematics_sample}
\begin{tabular}{llccccl}
\hline
ID & Alias & $z$ & Opt & $\beta_{\rm max}$ & $e_{\beta_{\rm max}}$ & Ref \\
\hline
0003+380 & S4 0003+38 & 0.229 & Q & 4.61 & 0.36 & \cite{2021ApJ...923...67H} \\
0003-066 & NRAO 005 & 0.3467 & B & 7.08 & 0.21 & \cite{2021ApJ...923...67H} \\
0010+405 & 4C +40.01 & 0.256 & Q & 6.92 & 0.64 & \cite{2021ApJ...923...67H} \\
0011+189 & RGB J0013+191 & 0.477 & B & 4.54 & 0.46 & \cite{2021ApJ...923...67H} \\
0014+813 & S5 0014+813 & 3.382 & Q & 9.47 & 0.91 & \cite{2021ApJ...923...67H} \\
0016+731 & S5 0016+73 & 1.781 & Q & 7.64 & 0.32 & \cite{2021ApJ...923...67H}\\
0027+056 & PKS 0027+056 & 1.317 & Q & 1.45 & 0.38 &\cite{2021ApJ...923...67H}   \\
0035+413 & B3 0035+413 & 1.353 & Q & 7.4 & 0.31 & \cite{2021ApJ...923...67H}\\
0044+566 & GB6 J0047+5657 & 0.747 & Q & 0.86 & 0.13 & \cite{2021ApJ...923...67H}\\
0048-071 & OB -082 & 1.975 & Q & 10.79 & 0.85 & \cite{2021ApJ...923...67H} \\
\hline
\end{tabular}\\
{\raggedright\footnotesize Note. Column (1) is the source ID; column (2) is the alias; column (3) is the redshift; column (4) is the optical class (Q = quasar, B = BL Lac, I = FR I, II = FR II); column (5) is the maximum apparent speed; column (6) is the uncertainty of the maximum apparent speed; column (7) is the reference. This table is available in its entirety in machine-readable form in the online supplementary material.\par}
\end{table}

\section{Results} \label{sec:static-results}

The distributions of the maximal proper motions for quasars/FR\,IIs and BL Lacs/FR\,Is are shown in Figure~\ref{fig:beta_Max_b}, while the corresponding distributions of the minimal Lorentz factors are shown in Figure~\ref{fig:gamma_Min_b}. The resulting minimal Lorentz factors range from $\sim$1.0 to 41.5 for quasars/FR\,IIs (average 11.5, median 9.5) and from $\sim$1.0 to 21.9 for BL Lacs/FR\,Is (average 4.2, median 1.5). We find that the observed maximal proper motions $\beta_{\rm app}$ of quasars are systematically higher than those of BL Lac objects, and the minimal bulk Lorentz factors of the fastest-moving jet components are accordingly higher for quasars than for BL Lacs. We note that only one BL Lac object has a proper motion higher than 20, i.e.,  $\beta_{\rm app}=21.9$ for 1823$+$568 (4C$+$56.27) with redshift $z=0.664$. Its host galaxy is very luminous, at the high end of the luminosity range of BL Lac objects \citep{1997ApJ...476..113F}. It was suggested to contain a binary BH, and then the ejection of VLBI components may be perturbed by the precession of the accretion disk and the motion of the black holes around the center of gravity of the binary black hole system \citep{2013A&A...557A..85R}. The detailed comparison of this source with the remaining BL Lacs is beyond the scope of this work. In summary, we find that around one tenth (6 of 56) of BL Lacs and FRIs have measured proper motions $\beta_{\rm app}>10$, while about one half (131 of 275) of quasars and FRIIs have $\beta_{\rm app}>10$. 	

It seems that the jets in BHXRBs are more similar to those in BL Lacs than to those in quasars, though a small fraction of BL Lacs have $\beta_{\rm app}>10$, which is different from the BHXRBs. Such discrepancy will be reconciled if the selection effects indeed play some roles in BHXRB observations as suggested by \citet{2026MNRAS.545f2102L}. The jets appear only in the low/hard or intermediate states in XRBs, where the accretion rates are lower than in the soft state. Similarly, it is believed that the accretion rates of BL Lacs/FR\,Is may be lower than those of quasars/FR\,IIs, implying different accretion modes between these two types of AGNs. We conjecture that the systematic difference in jet speeds may be regulated by the accretion modes. To interpret the statistical results, we develop a toy model of jet formation with different accretion modes in the following section.

\begin{figure}
	\centering
	\includegraphics[width=1.0 \columnwidth]{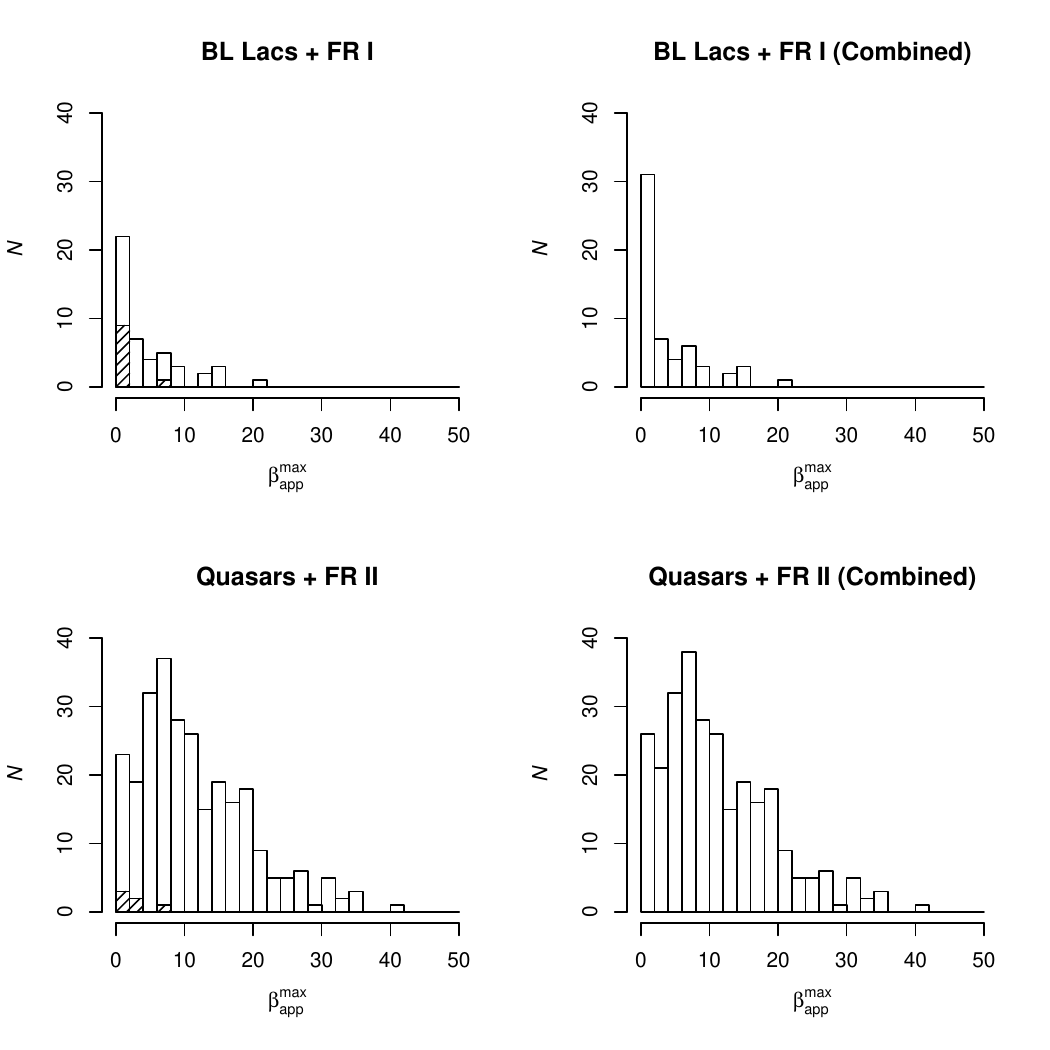}
	\caption{Distributions of maximal apparent speeds $\beta_{\mathrm{app}}^{\max}$ for different AGN types: (top-left) BL Lacs overlaid with FR I galaxies, (top-right) combined BL Lacs+FR I, (bottom-left) quasars overlaid with FR II galaxies, (bottom-right) combined quasars+FR II. Unfilled histograms represent the primary AGN types, and hatched histograms represent radio galaxies.} 
	\label{fig:beta_Max_b}
\end{figure}

\begin{figure}
	\centering
	\includegraphics[width=1.0 \columnwidth]{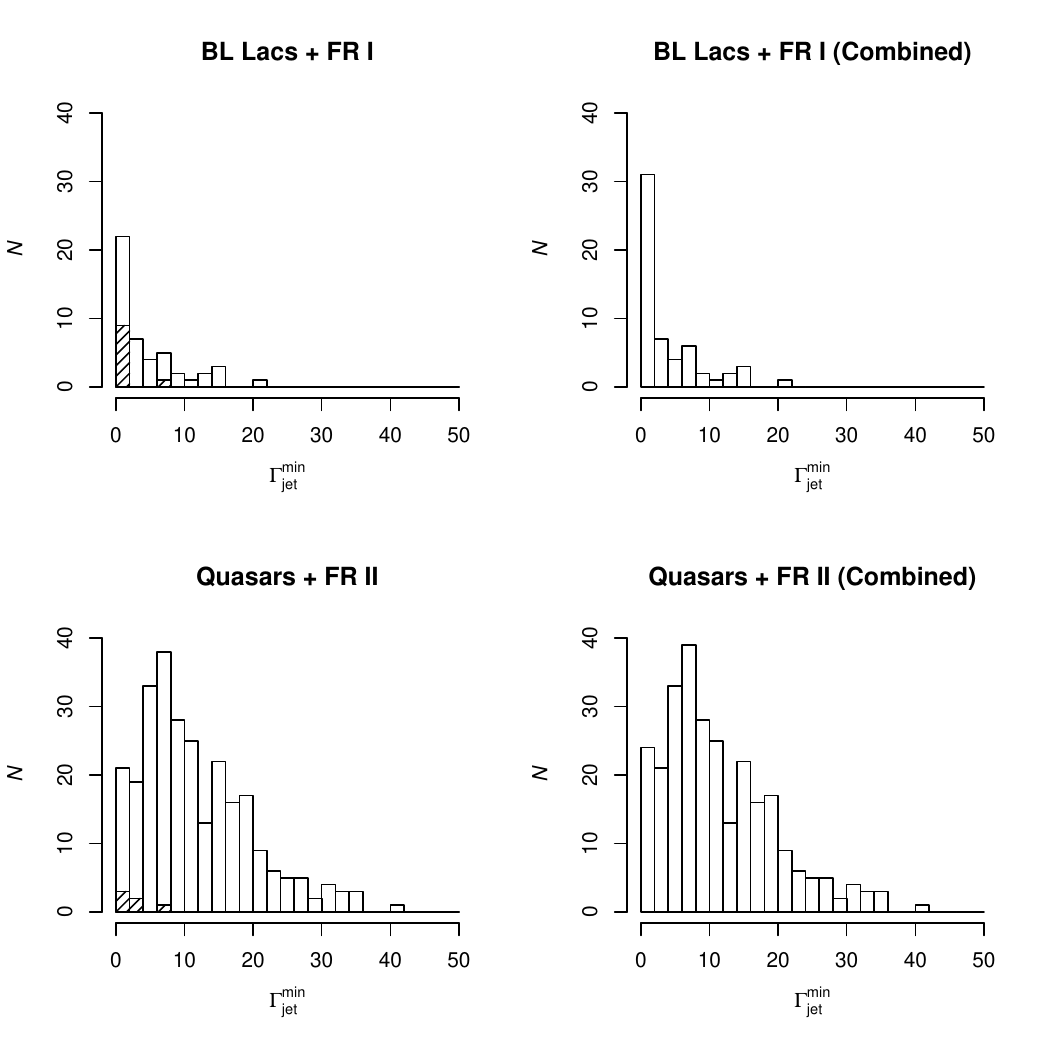}
	\caption{The same as Figure \ref{fig:beta_Max_b}, but for the distributions of minimal bulk Lorentz factors $\Gamma_{\mathrm{jet}}^{\min}$ derived from $\beta_{\mathrm{app}}^{\max}$. }	
    \label{fig:gamma_Min_b}
\end{figure}

\section{Magnetically driven jets} \label{sec:mag-jets}

It is widely believed that relativistic jets are accelerated by a rotating BH via the Blandford-Znajek (BZ) mechanism \citep{1977MNRAS.179..433B}. However, the field is thought to be sustained by currents in the surrounding disc \citep{1999ApJ...512..100L}. Although electron-positron pair plasma is usually assumed for BZ jets, it is likely that (part of) the jets are actually fed by gas plunging from the disc, co-rotating with the BH horizon frame \citep[see][for a detailed discussion]{2010LNP...794..233S, 2011MmSAI..82...95K, 2015ASSL..414...45T}. These features are indeed shown in numerical simulations \citep{2004ApJ...611..977M, 2005ApJ...620..878D}. A recent study shows that pair production through the BZ mechanism is orders of magnitude too low to explain the observed synchrotron emission of luminous jets in AGNs, strongly implying that matter loading of the jets from the surrounding medium is needed \citep{2026arXiv260200313M}. These two ideas are not mutually exclusive.

The dynamics of a magnetically accelerated jet can be described by the Bernoulli equation, derived from the conservation of mass and energy in the jet \citep[see][]{2007MNRAS.380...51K, 2011MmSAI..82...95K, 2015ASSL..414...45T, 2015MNRAS.451..927Z}. Detailed calculations of jet acceleration are challenging and beyond the scope of this paper. Assuming that all magnetic field energy at the jet base is converted into the kinetic energy of the jet's bulk motion, the jet's maximum speed can be estimated with
\begin{equation}
    \Gamma_{\rm jet}^{\rm max}=1+\sigma_0, \label{gamma_jet}
\end{equation}
where the magnetization parameter at the jet base is 
\begin{equation}
    \sigma_0\simeq {\frac {B_{\rm j,0}^2}{4\pi\rho_{\rm j,0}c^2}}, \label{sigma_0}
\end{equation}
in which the gas's internal energy is neglected relative to the rest energy of the mass, and $\Gamma_{\rm j,0}\sim 1$ is assumed at the base of the jet \citep{2011MmSAI..82...95K,2015ASSL..414...45T}.  

The relativistic jet is assumed to be driven from a region very close to the horizon of a rapidly spinning BH. The jet can be accelerated to a large Lorentz factor only if the magnetic field is sufficiently strong, the mass load rate is sufficiently low (i.e., $\rho_{\rm j,0}$ is low if the initial gas velocity at the jet base is similar), or both (see Equations \ref{gamma_jet} and \ref{sigma_0}). Otherwise, the magnetic field will drive a slow-moving outflow.

Observational evidence indicates that jet formation is closely related to hot gas in the accretion discs of both XRBs and AGNs \citep{2011MNRAS.416.1324Z, 2013ApJ...770...31W}. Magnetohydrodynamic calculations of magnetically launched outflows indicate that hot gas (probably in the corona) is necessary to launch an outflow from a thin disc, providing a natural explanation for the observational evidence that relativistic jets are associated with hot plasma in XRBs and AGNs \citep{2014ApJ...783...51C}. We assume that the jets are formed from hot gas (corona or ADAF) that is accelerated by the magnetic field co-rotating with the disc from a region very close to the BH horizon. Only a small fraction of the accreting gas is channeled into the jet from the region near the BH horizon, implying that the jet speed is governed by the properties of the accretion disc. For BL Lacs/FRIs or the low/hard state of the BHXRBs, the magnetic field is maintained by the ADAF, and the tenuous hot gas is magnetically driven from the upper layer of the ADAF. Thus, the field strength and gas density at the jet base are determined by the properties of the ADAF when the values of the parameters, $\alpha$ and $\dot{m}$, are specified. In the thin disc-corona case for quasars/FRIIs, the situation is more complicated than in the ADAF case. The field is maintained by the cold disc, and therefore the field strength at the jet base is determined by the disc properties that depend on $\alpha$ and $\dot{m}$, while the gas density at the jet base is related to the density of the corona above the disc. We summarize jet formation for the disc-corona system and the ADAF in the following two subsections, respectively. 

\subsection{Disc-corona} \label{sec:disk-corona}

In the disc-corona scenario, a small fraction of soft photons emitted from the disc are Compton upscattered to higher energies by hot electrons in the optically thin corona, producing a power-law hard X-ray continuum spectrum observed in quasars \citep{1979ApJ...229..318G, 1991ApJ...380L..51H}. The corona's vertical optical depth due to electron scattering, $\tau_{\rm es}\sim 0.1$--0.5, can be well constrained by the observed X-ray spectrum using disc-corona model calculations \citep[e.g.,][]{2009MNRAS.394..207C}. 

Calculations of the accretion flow surrounding a BH in the general relativistic (GR) frame show that the surface density of the accretion flow changes little as the gas plunges from the ISCO to the BH horizon \citep{2000ApJ...534..734M}. For a non-rotating BH, the surface density decreases by less than $\sim 50\%$, whereas for a rapidly spinning BH with $a=0.95$ it even increases slightly \citep[see Figure 2 in][]{2000ApJ...534..734M}. Motivated by these results, we assume that the corona's optical depth changes little as it plunges from the ISCO. The corona's density at the BH horizon is then estimated with
\begin{equation}
  \rho_{\rm h}^{\rm cor}\approx {\frac {\tau_{\rm es}}{H_{\rm cor}\kappa_{\rm T}}}=1.69\times 10^{-5}m^{-1}r_{\rm h}^{-1}\tilde{H}_{\rm cor}^{-1}\tau_{\rm es} ~{\rm g~cm^{-3}}, \label{rho_h_cor}
\end{equation}
where $\rho_{\rm h}^{\rm cor}$ is the mean corona density, $\kappa_{\rm T}$ is the electron-scattering opacity, $r_{\rm h}=R_{\rm h}c^2/GM$ ($R_{\rm h}$ is the BH horizon radius and $M$ is the BH mass), $m=M/M_\odot$, $H_{\rm cor}$ is the corona's scale height near $R\sim R_{\rm h}$, and $\tilde{H}_{\rm cor}=H_{\rm cor}/R$. 

The gas in the jet is supplied by the corona above the disc, so the gas density at the jet base should be related to the corona's density. We use a parameter $f_{\rm s}$ to estimate the gas density at the base of the jet, $\rho_{\rm j,0}=f_{\rm s}\rho_{\rm h}^{\rm cor}$, with $f_{\rm s}\lesssim 1$. Equation (\ref{sigma_0}) can be rewritten as
\begin{equation}
    \sigma_0^{\rm cor}\simeq {\frac {B_{\rm j,0}^2}{4\pi f_{\rm s}\rho_{\rm h}^{\rm cor}c^2}}. \label{sigma_0_cor}
\end{equation}

The detailed calculation of the magnetic field strength at the jet base in the GR frame is both challenging and model-dependent, so we assume it is related to the disc's pressure.
\begin{equation}
B_{\rm h}^{\rm disc}\sim 3.82\times 10^8\alpha^{-1/2}m^{-1/2}r_{\rm h}^{-3/4}\beta_{\rm b}^{-1/2}~{\rm G}, \label{b_h_cor}
\end{equation}
where $\beta_{\rm b}$ is the ratio of radiation to magnetic pressure in the disc, and an $\alpha$-viscosity is adopted \citep{2021aap...654A..81C}. We note that this estimate may be overly simplistic; however, the uncertainty can be absorbed into the parameter $\beta_{\rm b}$. Furthermore, we intend to compare the systematic differences between the corona and ADAF cases in this paper. These differences may be little affected by uncertainty if the two cases share similar physical processes.   

Substituting Equations (\ref{rho_h_cor}) and (\ref{b_h_cor}) into Equations (\ref{sigma_0_cor}), we have
\begin{equation}
\sigma_0^{\rm cor}\sim 0.76 \alpha^{-1} r_{\rm h}^{-1/2}\tilde{H}_{\rm cor}\tau_{\rm es}^{-1}  \beta_{\rm b}^{-1}f_{\rm s}^{-1}, \label{sigma_0_cor2}
\end{equation}
where $B_{\rm j,0}=B_{\rm h}^{\rm disc}$ are adopted. The maximal Lorentz factor of the jet can then be calculated using Equation (\ref{gamma_jet}). 

\subsection{ADAF}  \label{sec:adaf}

As in the thin accretion disc-corona case, we estimate the field strength and the density of an ADAF at the BH horizon,
\begin{equation}
B_{\rm h}^{\rm ADAF}=6.73 \times 10^8 \alpha^{-1/2}m^{-1/2}\dot{m}^{1/2}r_{\rm h}^{-5/4}\beta_{\rm b}^{-1/2}~{\rm G}, \label{b_h_adaf}
\end{equation}
and
\begin{equation}
    \rho_{\rm h}^{\rm ADAF}=9.99\times 10^{-5} \alpha^{-1}m^{-1}\dot{m}r_{\rm h}^{-3/2} ~{\rm g~cm^{-3}}, \label{rho_h_adaf}
\end{equation}
where the self-similar solution is adopted and $\dot{m}$ is the accretion rate in units of the Eddington rate \citep[see][for details]{1995ApJ...452..710N}. Substituting Equations (\ref{b_h_adaf})-(\ref{rho_h_adaf}), along with $\rho_{\rm j,0}=f_{\rm s}\rho_{\rm h}^{\rm ADAF}$ and $B_{\rm j,0}=B_{\rm h}^{\rm ADAF}$, into Equation (\ref{sigma_0}), we obtain the magnetization parameter at the jet base,
\begin{equation}
\sigma_0^{\rm ADAF}\sim 0.40 r_{\rm h}^{-1}\beta_{\rm b}^{-1}f_{\rm s}^{-1}, \label{sigma_0_adaf2}
\end{equation}
with which the maximal Lorentz factor of the jet is given by Equation (\ref{gamma_jet}).

\subsection{Comparison between two accretion modes} \label{sec:comparison}

We note that our results are independent of the accretion rate in both accretion modes. Although the precise value of the model parameter $f_{\rm s}$ is uncertain and may vary across sources, the systematic difference derived in our model calculations between the accretion disc-corona and the ADAF cases should be robust, because the gas dynamics are nearly the same in both cases. This can be clarified by the ratio of the magnetization factors for these two cases,
\begin{equation}
   {\frac {\sigma_0^{\rm cor}}{\sigma_0^{\rm ADAF}}}=1.90\alpha^{-1}r_{\rm h}^{1/2}\tilde{H}_{\rm cor}\tau_{\rm es}^{-1}. \label{ratio_sigma}
\end{equation}
The ratios of magnetic field strength and density between the disc-corona and ADAF cases are
\begin{equation}
    {\frac {B_{\rm h}^{\rm disc}}{B_{\rm h}^{\rm ADAF}}}=0.57 \dot{m}^{-1/2}r_{\rm h}^{1/2}, \label{ratio_b}
\end{equation}
and 
\begin{equation}
    {\frac {\rho_{\rm h}^{\rm cor}}{\rho_{\rm h}^{\rm ADAF}}}=0.17\alpha\dot{m}^{-1}r_{\rm h}^{1/2}\tilde{H}_{\rm cor}^{-1}\tau_{\rm es}, \label{ratio_rho}
\end{equation}
respectively. The division of the Eddington ratio between BL Lac objects/FR Is and quasars/FR IIs is $\sim 0.01$ \citep{2001aap...379L...1G, 2009ApJ...694L.107X}, roughly consistent with the critical accretion rate $\dot{m}_{\rm crit}\sim 0.01$ for an ADAF with $\alpha=0.1$ \citep[see][for details]{1998tbha.conf..148N}. Thus, ${B_{\rm h}^{\rm disc}}\gtrsim 5{B_{\rm h}^{\rm ADAF}}$ and $\rho_{\rm h}^{\rm cor}\simeq \rho_{\rm h}^{\rm ADAF}$ for an ADAF accreting at the critical rate around a rapidly spinning BH, assuming typical corona parameters $\tau_{\rm es}=0.3$ and $\tilde{H}_{\rm cor}=0.5$. In this case, the magnetization factor $\sigma_0^{\rm cor}\gtrsim 25\sigma_0^{\rm ADAF}$ (cf. Equation \ref{sigma_0}). We note that $\rho_{\rm h}^{\rm ADAF}\propto \dot{m}$ and $B_{\rm h}^{\rm ADAF}\propto \dot{m}^{1/2}$, implying that the magnetization factor $\sigma_0^{\rm ADAF}$ is independent of $\dot{m}$ (see Equation \ref{sigma_0_adaf2}). A larger BH spin parameter $a$ corresponds to a smaller $r_{\rm h}$, which leads to a stronger magnetic field $B_{\rm j,0}$ at the jet base and, consequently, a larger Lorentz factor. For a rapidly spinning BH with $a=0.95$, we have ${\sigma_0^{\rm cor}}/{\sigma_0^{\rm ADAF}}=3.63\alpha^{-1}$ for $\tau_{\rm es}=0.3$ and $\tilde{H}_{\rm cor}=0.5$ (see Equation \ref{ratio_sigma}). Even if an upper limit on the viscosity parameter $\alpha=1$ is adopted, the Lorentz factor of the jet driven from the corona of the disc is still substantially larger than that of the jet from the ADAF. 

\begin{figure}
	\centering
	\includegraphics[width=0.8 \columnwidth]{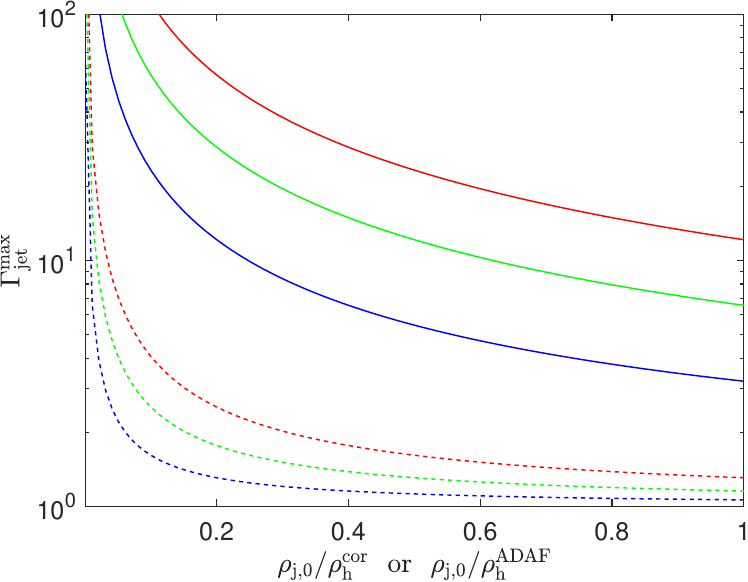}
	\caption{The bulk Lorentz factors of the jets magnetically accelerated from the BH horizon depend on the density ratio of the jet base to the ADAF or the corona, i.e., the dimensionless mass-loss rate in the jets (see the text in Section \ref{sec:disk-corona}). The solid lines show the disc-corona case calculated with Equations (\ref{gamma_jet}) and (\ref{sigma_0_cor2}), while the dashed lines show the ADAF case calculated with Equations (\ref{gamma_jet}) and (\ref{sigma_0_adaf2}). The colored lines indicate results for different values of $\beta_{\rm b}$:  $\beta_{\rm b}=1$ (red), 2 (green), and 5 (blue). In all our calculations, we adopt the viscosity parameter $\alpha=0.1$, the BH spin parameter $a=0.95$, the optical depth of the corona $\tau_{\rm es}=0.3$, and the relative thickness of the corona $\tilde{H}_{\rm cor}=0.5$.}
	\label{fig:gamma_jet}
\end{figure}

\section{Discussion} \label{sec:discussion}

It is well known that the accretion mode transitions as the accretion rate approaches a critical value, below which an ADAF is present in the inner region of the disc \citep{1994ApJ...428L..13N, 1995ApJ...452..710N}. Otherwise, a disc with a hot corona appears when the accretion rate exceeds the critical value \citep{1979ApJ...229..318G, 1991ApJ...380L..51H}. The Lorentz factor varies with the BH spin parameter $a$ through the BH horizon radius $r_{\rm h}$.

The calculations of a jet magnetically accelerated from a rotating BH in the general relativistic frame are beyond the scope of this work. Instead, we aim to derive the basic properties of a relativistic jet using an approach similar to that described in \citet{2011MmSAI..82...95K}. The terminal speed of the jet accelerated by a rotating magnetic field is governed by the dimensionless magnetization parameter, which depends on the jet density (mass-loss rate) and the magnetic field strength \citep[see][for a detailed discussion]{2011MmSAI..82...95K, 2015ASSL..414...45T}. In the calculations of jet acceleration, we consider two cases: an accretion disc with a hot corona accreting at a high rate, and an ADAF when the accretion rate is low (see Section \ref{sec:mag-jets} for details). The results are plotted in Figure \ref{fig:gamma_jet}, which shows that jet speeds are substantially higher for BHs surrounded by accretion discs with coronas than for those with ADAFs. For the disc-corona case, both the magnetic field strength and the corona's density are independent of the accretion rate, whereas for the ADAF case $B_{\rm ADAF}\propto \dot{m}^{1/2}$ and $\rho_{\rm ADAF}\propto \dot{m}$. Thus, the results for jet speed are independent of BH mass or accretion rate in both the disc-corona and ADAF cases (see Equation \ref{sigma_0}). The magnetic field strength ratio of disc to ADAF increases with the ADAF accretion rate, and we find that the disc's field is substantially stronger than that of the ADAF even if the ADAF is accreting at the critical rate, whereas its density is roughly equal to the corona's density (see Section \ref{sec:comparison}), resulting in much faster jet speeds for the disc-corona case than for the ADAF case. The fields of the jet components in radio-loud quasars estimated with a jet emission model from their SEDs are systematically stronger than those in BL Lac objects \citep{2023ApJS..268...23F}. It was suggested that the field strength of the jet component is closely related to that near the BH horizon \citep{2014Natur.510..126Z}. Therefore, their results may imply different field strengths in these two types of the sources, which is qualitatively consistent with our conclusion derived in this work.      

It is generally accepted that the physics of accretion disks and jets should be the same for BHXRBs and AGNs. Our investigation shows that the Lorentz factors of the jets in BL Lac objects are similar to those of BHXRBs. Jet speeds are governed by the magnetic field strength and the mass-loading rate. For BHXRBs, relativistic jets are present only in the low/intermediate states, when hot accretion flows surround the BHs. The magnetic field is limited by the gas pressure of the hot accretion flow. Because the accretion rate is low, the field near the BH horizon remains relatively weak, which usually leads to a mildly relativistic jet.  

Jets are suppressed in the high states of BHXRBs, whereas they are ubiquitously observed in radio quasars, a pattern explained by the mass- and field-flux feeding scenario proposed by \citet{2019MNRAS.485.1916C}. The bulk Lorentz factors of jet components in quasars are substantially higher than those of BL Lac objects, giving the impression that AGN jet speeds are much higher than those of BHXRBs. Our calculations show that jets accelerated from the hot gas above the disc near the BH horizons are indeed much faster than those in ADAFs, since the field maintained by a thin disc is significantly stronger than that in the ADAF (see the discussion in Section \ref{sec:adaf}). We note that a small fraction of BL Lacs show relatively high jet speeds with $\beta_{\rm app}>10$. One possibility is that the density ratios of the jet base to ADAF are much lower than in other cases. The Lorentz factors of the jets can be up to $\sim 20$ if the density ratio is as low as $0.01$ and $\beta_{\rm b}=1$ is adopted (see Figure \ref{fig:gamma_jet} for ADAF cases). The values of the density ratios of the jet base to ADAF are sensitive to the field configuration \citep[see][for the discussion]{1994A&A...287...80C,2002A&A...385..289C,2010LNP...794..233S}, the details of which are beyond the scope of this work. The field strengths of the ADAFs in these BL Lacs are probably stronger than those of other BL Lacs with relatively lower proper motions. The ADAFs in these sources may be in a magnetically arrested state \citep{2003PASJ...55L..69N,2022MNRAS.511.3795N}, and therefore we suggest that they are suitable candidates for exploring the physics of magnetically arrested discs in the future.  

We note that this work considers only steady jets in XRBs, which are the same type of jets observed in AGNs. The transient ejecta in XRBs should have a very different nature, owing to their large inertia, which allows them to reach pc scales. It has been suggested that powerful transient jets in XRBs may be related to BH spins \citep[see the discussion in][]{2026NewAR.10201746Z}. It remains unclear whether such transient jets are present in AGNs, and our model is not applicable to them. 

\section*{Acknowledgements}

We thank the anonymous referee for constructive comments and suggestions that have improved the clarity and quality of this manuscript. This work is supported by the NSFC (12533005, 12233007, and 12347103), the science research grants from the China Manned Space Project with CMS-CSST-2025-A07, and the fundamental research funds for Chinese central universities (Zhejiang University). AAZ acknowledges support from the Polish National Science Center grant 2023/48/Q/ST9/00138.

\section*{Data Availability}
The full sample table (333 sources, including ID, alias, redshift, optical class, maximum apparent speed, its uncertainty, and reference) is available at the CDS VizieR service (https://cdsarc.cds.unistra.fr/). 
The data used to compile this sample are primarily drawn from the MOJAVE program and supplemented with additional literature sources as described in Section~2. The codes used for the analysis are available upon request to the corresponding authors.



\bibliographystyle{mnras}
\bibliography{xrb_jet-speed-bibliography}








\bsp	
\label{lastpage}
\end{document}